\documentclass[12pt]{article}
\usepackage{amsfonts}
\usepackage{amssymb}
\usepackage{graphicx}
\newcounter{rown}

\def\siz{\small}

\begin{document}
\title{EXOTIC GALILEAN CONFORMAL  SYMMETRY AND ITS
DYNAMICAL REALISATIONS}
\author{J. Lukierski$^{1)}$, P.C. Stichel$^{2)}$  and W.J. Zakrzewski$^{3)}$
\\
\siz $^{1)}$Institute for Theoretical Physics,  University of
Wroc{\l}aw, \\ \siz pl. Maxa Borna 9,
 50--205 Wroc{\l}aw, Poland\\
 \siz e-mail: lukier@ift.uni.wroc.pl\\
\\\siz
$^{2)}$An der Krebskuhle 21, D-33619 Bielefeld, Germany \\ \siz
e-mail:peter@physik.uni-bielefeld.de
\\ \\ \siz
$^{3)}$Department of Mathematical Sciences, University of Durham, \\
\siz Durham DH1 3LE, UK \\ \siz
 e-mail: W.J.Zakrzewski@durham.ac.uk
 }

\date{}
\maketitle

\begin{abstract}
The six-dimensional exotic Galilean algebra in (2+1) dimensions with two central charges
$m$ and $\theta$,  is extended when $m=0$, to a ten-dimensional
Galilean   conformal algebra with dilatation, expansion, two
acceleration generators and the central charge $\theta$.
A realisation of such a symmetry is provided by a model
with higher derivatives recently discussed in \cite{peterwojtek}.
We consider also a realisation of the Galilean conformal
symmetry for the  motion with a Coulomb
potential and a magnetic vortex interaction. Finally, we study the restriction, as well as
the modification,
of the Galilean conformal algebra obtained after the introduction of the
minimally coupled  constant electric and magnetic fields.
\end{abstract}
\section{Introduction}

There are two possible ways of looking at symmetries in physics. One of them asks which symmetry is possessed
by a given model, {\it i.e.} one tries to find a specific realisation of the symmetry generators in a model and  then calculate
their Lie algebra. The second method involves going the other way round, {\it i.e.}
one looks for a concrete realisation of a given symmetry algebra by constructing new models.

In 1997 we followed the second approach and presented a Lagrangian
point particle model which possessed planar Galilean symmetry with
two central charges $m$ and $\theta$. To construct such a model we had to
introduce a higher-order Lagrangian \cite{lsz}
\begin{equation}\label{oneone}
L\,=\,\frac{m}{2}\dot x_i^2\,-\,\frac{\theta}{2}\,\epsilon_{ij}\,\dot x_i\,\ddot x_j,
\end{equation}
where $\theta$ corresponded to the second central charge appearing in the
Poisson bracket of the two boost generators $K_i$
\begin{equation}\label{onetwo}
\{K_i,\,K_j\}\,=\,\theta\,\epsilon_{ij}.
\end{equation}
Then, in \cite{peterwojtek} the $m=0$ limit of (\ref{oneone}) was considered.

There are two ways of extending the Galilean symmetry:
\begin{itemize}
\item i) One can add the dilatation and conformal transformations
preserving the Schr\"odinger equation. In this approach one adds
to the Galilean algebra, in any spatial dimensions, two additional
generators: dilatation $D$ and expansion $K$. The resultant
algebra is called the Schr\"odinger algebra [3-5]. However,
although this symmetry was also called `nonrelativistic conformal
symmetry' in [3] it does not inherit the basic properties of
relativistic conformal symmetries (vanishing of the mass
parameter, the number of conformal generators being equal to the
number of translations etc.) \item ii) One can perform the
nonrelativistic contraction of the relativistic conformal algebra
in $D$ dimensions isomorphic to $O(D,2)$ algebra.
 We supplement the Poincar\'{e} algebra generators ($P_{\mu},$ $J_{\mu\nu}$), ($\mu,\nu=0,1,...D-1$)
by dilatation generators $D$ and special conformal generators $R_{\mu}$. We rescale
the generators in the following way: ($i=1,...D-1)$
\begin{equation}\label{a}
P_0\,=\,\frac{H}{c},
\end{equation}
\begin{equation}\label{b}
J_{i0}\,=\,cK_i,
\end{equation}
\begin{equation}\label{c}
R_i\,=\,c^2F_i,\qquad R_0\,=\,cK.
\end{equation}
The remaining generators $P_i$, $J_{ij}$ and $D$ remain unscaled.
\end{itemize} We see that
\begin{itemize}
\item i) The relation (3) implies that we should put the rest mass $m_0=0$ (in general we have the expansion $P_0=m_0c+\frac{H}{c}$).
\item ii) With the rescaling (3-5) the contraction limit $c\rightarrow\infty$ does exist and describes
a proper non-relativistic conformal extension
of Galilean symmetries. It should be added that Negro et al
introduced  in \cite{Negro} a family of nonrelativistic conformal algebras dependent
on half-integer $l$. When $l=1$ one obtains
 a nonrelativistic conformal algebra, which
coincides with the one described by the
$c\rightarrow \infty$ limit. We shall call this algebra
the Galilean conformal algebra, with 10 generators in $D=2+1$ and
$15$ generators in $D=3+1$.
\end{itemize}

The aim of this paper is to study the symmetries of the nonrelativistic conformal models
in (2+1) dimensions, in the presence of the central extension $\theta$.

The paper is organised as follows. In the next section we perform the
contraction of the $D=3$ relativistic conformal algebra (isomorphic to $O(3,2)$)
 and show that by adding the central charge $\theta$ we derive the exotic (2+1) dimensional
Galilean conformal algebra.
In section 3 we demonstrate that
the free model introduced in \cite{peterwojtek}
is a realisation of the  exotic Galilean conformal symmetry.
We discuss also how the Galilean conformal symmetry is modified if the
Coulomb term and/or magnetic vortex interactions are added.  Finally, in section 4,
we show, by extending the results of [7], how the Galilean symmetry
group is changed when we add to the model of \cite{peterwojtek}
 constant electromagnetic fields.

\section{ Exotic Galilean conformal group in (2+1)
 dimensions}

We define the $D=3$ relativistic conformal algebra by adding to
the $D=3$ Poincar\'{e} algebra the following nonvanishing
commutators ($\mu,\nu,\rho=0,1,2$; $i,j=1,2$:
$\eta_{\mu\nu}=\mbox{diag}(1,-1,-1))$ (see {\it e.g.} [8])
\begin{equation}\label{dd}
[J_{\mu\nu},\,R_{\rho}]\,=\,\eta_{\nu\rho}R_{\mu}\,-\,\eta_{\mu\rho}R_{\nu},
\end{equation}
\begin{equation}\label{ddd}
[R_{\mu},\,P_{\nu}]\,=\,-2(\eta_{\mu\nu}D\,+\,J_{\mu\nu}),
\end{equation}
\begin{equation}\label{dddd}
[D,\,P_{\mu}]\,=\,-P_{\mu},\quad [D,\,R_{\mu}]\,=\,R_{\mu}.
\end{equation}
We introduce $J_{\mu\nu}=(J_{ij}=\epsilon_{ij}J,\,J_{i0}=cK_i)$,
$P_{\mu}=(P_i,\frac{H}{c}),$ $R_{\mu}=(c^2F_i,cK)$ and keep $D$
nonscaled. By  performing the nonrelativistic $c\rightarrow\infty$
limit
 we obtain the 10-dimensional Lie-algebra,
with generators $P_i$ (space translations), $K_i$ (Galilean boosts), $J$ ($O(2)$ rotations),
$H$ (time translations), $D$ (dilatations), $K$ (time expansions) and $F_i$ (accelerations).
We list below only the nonvanishing commutators.
First, for any vector $A_i\in (P_i,\,K_i,\,F_i)$
we have the following Lie-bracket with respect to rotations
\begin{equation}\label{twoone}
[J,\,A_{i}]\,=\,\epsilon_{ij}\,A_{j}.
\end{equation}
Furthermore
 \begin{equation}\label{twotwo}
[H,\,K_{i}]\,=\,P_{i},\qquad [H,\,F_i]\,=\,2K_i.
\end{equation}
The one-dimensional conformal subalgebra (see {\it e.g.} [9])
is the following:
\begin{equation}\label{twotheree}
[D,\,H]\,=\,-H,\quad [K,\,H]\,=\,-2D,\,\quad [D,\,K]\,=\,K.
\end{equation}
We have further
\begin{equation}\label{twofive}
[D,\,P_{i}]\,=\,-\,P_{i},\qquad [D,\,K_i]\,=\,0,\qquad [D,\,F_i]\,=\,F_i
\end{equation}
and finally
\begin{equation}\label{twofivea}
[K,\,P_{i}]\,=\,-2\,K_{i},\qquad [K,\,K_i]\,=\,-\,F_i,\qquad [K,\,F_i]\,=\,0.
\end{equation}

The realisation of the Lie algebra  (9-13) on the $D=(2+1)$
nonrelativistic  space and time, which can be also obtained by the
contraction $c\rightarrow \infty$ of the space-time differential
realisation of the $D=3$ relativistic conformal algebra (see {\it
e.g.} [8]), is, after putting $x_0=ct$, given in terms of
differential operators: (see also \cite{Negro}):
$$ H\,=\,\partial_t,\qquad P_{i}\,=\,-\partial_i,\qquad K_i\,=\,-t\partial_i,\qquad F_i\,=\,-t^2\partial_i,$$
\begin{equation}\label{twosix}
J\,=\,-\,\epsilon_{ij}\,x_i\,\partial_j,
\end{equation}
$$D\,=\,t\partial_t\,+\,x_i\partial_i,\qquad K\,=\,t^2\partial_t
\,+\,2t\,x_i\partial_i.$$

The Galilean conformal algebra  in (2+1) dimensions can have a central
extension by an `exotic' parameter $\theta$. This parameter is introduced into
 the Lie bracket for two Galilean boosts (see \cite{lsz}):
\begin{equation}\label{twoten}
[K_i,\,K_j]\,=\,\theta\,\epsilon_{ij}.
\end{equation}

As a consequence the Lie-bracket $[P_i, F_j]$ becomes also nonvanishing:
\begin{equation}\label{twoeleven}
 [P_i,\,F_j]\,=\,-2\,\theta\,\epsilon_{ij}.
 \end{equation}

\noindent  To prove (\ref{twoeleven}) we use the relation
 $F_j\,=\,-[K,\,K_j]$
 which gives, by the Jacobi identity,
 \begin{equation}\label{twothirteen}
 [P_i,\,F_j]\,=\,[K,\,[K_j,\,P_i]]\,+\,[K_j,\,[P_i,\,K]].
 \end{equation}

 \noindent However, because $[K_j,P_i]=0$ (as $m=0$),
 and using
 $[P_i,K]=2K_i$ as well as (\ref{twoten}), we see that the relation (\ref{twothirteen})
 leads to the formula (\ref{twoeleven}).

The Galilean conformal algebra with the modified relations (15), (16), in what follows, will be called
the `exotic Galilean conformal algebra'.

We shall see that in the model introduced in [1], the exotic Galilean conformal algebra is enlarged by
 two further generators $J_{\pm}$ extending
the $O(2)$ spatial rotations to the special linear group $sl(2)\sim O(2,1)$ (we put $J_3=\frac{J}{2},$
$J_{\pm}=J_1\pm iJ_2$, where $J_r\,\, (r=1,2,3)$ are the standard $O(2,1)$ generators).
\begin{equation}\label{twotwelelwe}
[J_3,\,J_{\pm}]\,=\,\mp iJ_{\pm},\quad [J_+,\,J_-]\,=\,2iJ_3.
\end{equation}
The remaining nonvanishing commutators of $J_{\pm}$ describe
the $sl(2)$ covariance relations for any two-vector $A_i\in (P_i,K_i,F_i)$,
\begin{equation}\label{twoeight}
 [J_+,\,A_{-}]\,=\,-i\,A_{+},\qquad [J_-,\,A_{+}]\,=\,i\,A_{-},
\end{equation}
where $A_{\pm}=A_1 \pm iA_2$.

The differential realisation of the generators $J_{\pm}$  is given by
\begin{equation}\label{twosevenc}
J_{\pm}\,=\,\pm\,i\,x_{\pm}\partial_{\mp},
\end{equation}
where $x_{\pm}=x_1\pm ix_2$ and $\partial_{\pm}=\frac{\partial}{\partial x_{\pm}}=\frac{1}{2}(\partial_{x_1}\mp i \partial_{x_2}).$

It is an open question whether the extension by the generators $J_{\pm}$ always exists
if we deal with Lagrangian models in $D=(2+1)$ which are invariant under the exotic Galilean
conformal symmetry.

\section{Galilean conformal symmetry ($D=(2+1))$ in dynamical models}

The planar model introduced in \cite{peterwojtek}, in the noninteracting case, is defined by the first order Lagrangian\footnote{cp. to \cite{peterwojtek} we have changed the sign of $\theta$ in accordance with
(\ref{twoten}).}
\begin{equation}\label{threeone}
L_{0}\,=\,P_i(\dot x_i-y_i)\,-\,\frac{\theta}{2}\,\epsilon_{ij}\,y_i\,\dot y_j.
\end{equation}
This expression is the $m\to 0$ limit of the model
introduced in \cite{lsz}.

The model described by (21) possesses the full exotic Galilean
conformal symmetry.
Indeed, the corresponding Lie algebra
in $D=(2+1)$ is realised as the
Poisson-bracket (PB) algebra expressed in terms
of phase space variables $x_i$, $y_i$ and $P_i$ as follows:
\begin{equation} \label{threethree}
H\,=\,P_iy_i,
\end{equation}
\begin{equation}\label{threefour}
P_i\,=\,P_i,
\end{equation}
\begin{equation}\label{threefive}
K_i\,=\,P_it\,+\,\theta\,\epsilon_{ij}\,y_j,
\end{equation}
\begin{equation}\label{threesix}
F_i\,=\,-t^2P_i\,+\,2tK_i\,-\,2\theta\,\epsilon_{ij}\,x_j,
\end{equation}
\begin{equation}\label{threeseven}
J\,=\,\epsilon_{ij}\,x_i\,P_j\,-\, \frac{\theta}{2}y_i^2,
\end{equation}
\begin{equation}\label{threeeight}
D\,=\,tH\,-\,x_iP_i,
\end{equation}
\begin{equation}\label{threenine}
K\,=\,-t^2H\,+\,2tD\,-\,2\theta\,\epsilon_{ij}\,x_iy_j.
\end{equation}

Let us observe that the second term in formula (\ref{oneone}) is invariant under the
$Sp(2)\sim O(2,1)\sim SL(2)$ transformations
\begin{equation}\label{threeninea}
x_i\,=\,A_{ij} x_k,\qquad A^{T}\epsilon A\,=\,\epsilon\
\end{equation}
{\it i.e.} the $O(2)$ generator $J$ should be supplemented by the
generators $J_{\pm}=J_1\pm iJ_2$ satisfying the $O(2,1)$ algebra (13).
In order to obtain the extended Galilean conformal invariance of the first order Lagrangian
(21) we should add to the transformation (29)
the relations:
\begin{equation}\label{new}
y_i'\,=\,A_{ij}y_j,\qquad P_i'\,=\,(A^{-1})_{ij}P_j.
\end{equation}
Moreover,  the generators $J_{\pm}$ have in the phase space of our model
the following realisation:
\begin{equation}\label{threeten}
J_{\pm}\,=\,-\frac{\theta}{4}\,y_{\pm}^2\,\mp\,\frac{i}{2}\,x_{\pm}P_{\pm}.
\end{equation}

 To prove that all the generators ((22-28) and (31)) are conserved we use the
equations of motion (EOM) derived from the Lagrangian $L_0$
\begin{equation}\label{threeeleven}
\dot x_i\,=\,y_i,\qquad \dot P_i\,=\,0,
\end{equation}
$$\dot y_i\,=\,\frac{1}{\theta}\,\epsilon_{ik}\,P_k.
$$

\noindent These equations can be written as Hamilton's equations
\begin{equation}\label{threetwelve}
\dot Y\,=\,\{Y,\,H\},\qquad Y\,\in\,(x_i,y_i,P_i)
\end{equation}
by using the following nonvanishing fundamental PBs

\begin{equation}\label{threetwelvea}
\{x_i,\,P_k\}\,  = \,\delta_{ik},\qquad
 \{y_i,\,y_k\}\,=  \,\frac{\epsilon_{ik}}{\theta}
\end{equation}
obtained from (21) by the canonical procedure due to Faddeev and Jackiw \cite{jackiw}.

By means of these PBs it is straightforward to show that the generators (\ref{threethree}-28) satisfy the exotic
Galilean conformal algebra
extended by
the generators $J_{\pm}$ (see section 2).

Let us consider now the model (21) as describing the motion
with the Coulomb and magnetic vortex interactions ($r=(x_1^2+x_2^2)^{\frac{1}{2}}$)
\begin{equation}\label{threetwo}
L_{int}\,=\,\frac{\lambda}{r}\,-\,\frac{g}{r^2}\,\epsilon_{ij}\,y_i\,x_j.
\end{equation}

In this case the variables $x_i$ and $y_i$ can be treated as
coordinate differences, {\it i.e.} as invariant under the translations, boosts and constant
accelerations. Therefore, we should consider the representation of the Galilean conformal algebra with six vanishing
generators
\begin{equation}\label{newa}
P_i\,=\,K_i\,=\,F_i\,=\,0.
\end{equation}
The four remaining generators ($H,D,K$ and $J$) form an $O(2,1)\oplus O(2)$
algebra described by the algebra (11) supplemented by the Abelian $O(2)$ rotation
generator. Because the second term in (35) is, like the free model, invariant
under $Sp(2)\sim O(2,1)$ transformations (29), when
$\lambda=0$ and $g\ne 0$ the model possesses the symmetry $O(2,1)\otimes O(2,1)$, described by the algebras
(11) and (18).

\section{Enlargement of the exotic Galilean symmetry in the presence of constant electromagnetic fields}

 In this section we couple minimally the model of \cite{peterwojtek} to the
 constant electromagnetic fields $B$ and $E_i$ and show how the exotic Galilean conformal algebra in $D=(2+1)$
is modified into an enlarged Galilean symmetry.
  In such a case we consider $E_i$ and the corresponding canonical conjugate momenta $\pi_i$ as additional phase space variables
 (cp \cite{ho}).

 The minimal coupling principle gives
 \begin{equation}\label{fourone}
 H_0\,\to\,H\,=\,H_0\vert_{P_i\rightarrow P_i-A_i}\,-\,A_0,
 \end{equation}
 which, for the constant fields $B$ and $E_i$, modifies
 $L_0$ (\ref{threeone}) in the following way
 \begin{equation}\label{fourtwo}
 L_0\,\rightarrow\, L\,=\,P_i\dot x_i\,-\,(P_i\,+\,\frac{B}{2}\epsilon_{ij}x_j)y_i\,-\,
 \frac{\theta}{2}\epsilon_{ij}y_i\dot y_j\,+\,E_ix_i\,+\,\dot{E_i}\pi_i.
 \end{equation}
 Note that (\ref{fourtwo}) leaves the PBs {34) unchanged and now they have to
supplemented by
\begin{equation}\label{fourthree}
\{E_i,\,\pi_j\}\,=\,\delta_{ij}.
\end{equation}

The EOM which follow from the Lagrangian (38) are given by
\begin{equation}\label{fourfour}
  \dot x_i\,=\,y_i,\qquad \dot \pi_i\,=\,x_i
\end{equation}
\begin{equation}\label{fourfive}
 \dot P_i\,=\,\frac{B}{2}
\epsilon_{ik}y_k\,+\,E_i,
\end{equation}
\begin{equation}\label{foursix}
\dot y_i\,=\,\frac{1}{\theta}
(\epsilon_{ik}P_k-\frac{B}{2}x_i)
\end{equation}
\begin{equation}\label{fourseven}
  \dot E_i\,=\,0.
\end{equation}

The EOM for $x_i$ follows from (40-43):
\begin{equation}\label{fournine}
\dot{\ddot{x_i}}\,=\,-\frac{B}{2}\,\dot x_i\,+\,\frac{\epsilon_{ik}}{\theta}E_k.
\end{equation}

>From (\ref{fournine}) we can read off the symmetries of the model.
They are given by:
\begin{itemize}
\item Space translations: $\delta x_i=a_i$ with $\delta B=\delta E_k=0$
\item Rotations: $\delta x_i=-\varphi \epsilon_{ik}x_k$ with $\delta B=0$
and $\delta E_i=-\varphi \epsilon_{ik}E_k.$
\item Boosts: $\delta x_i=b_i t$ with $\delta B=0$ and $\delta E_i=-B\epsilon_{ij}b_j$
\end{itemize}
corresponding to the physically required transformation properties
of the electromagnetic fields in the nonrelativistic limit.

In addition we have also the
additional $O(2,1)$ invariance with
the generators ($J_{\pm},J_3)$
which satisfy (18).
 To see this we rewrite (\ref{fournine}) in terms of the complex variables ($A_{\pm}=A_1\pm iA_2$ for any vector $A_i$) getting
\begin{equation}\label{fourninea}
\dot{\ddot{x_+}}\,=\,-\frac{B}{\theta}\dot x_+\,-\,\frac{i}{\theta}E_+
\end{equation}
\begin{equation}\label{fournineb}
\dot{\ddot{x_-}}\,=\,-\frac{B}{\theta}\dot x_-\,+\,\frac{i}{\theta}E_-
\end{equation}
and observe that
$\delta_+(\ref{fournineb})\,\rightarrow\, (\ref{fourninea})\,\mbox{and}\,
\delta_-(\ref{fourninea})\, \rightarrow\, (\ref{fournineb})$
where $\delta_{\pm}$ act on coordinates and electric fields as
\begin{equation}\label{aaaa}
 \delta_+x_-= -i\epsilon x_+,\qquad \delta_+E_-=i\epsilon E_+,\end{equation}
\begin{equation}\label{bbbb}
 \delta_-x_+= i\epsilon x_-,\qquad \delta_-E_+=-i\epsilon E_-\end{equation}
with all other variations being zero.

All symmetries extending conformally our exotic (2+1) dimensional Gali-
lean algebra,
{\it i.e.} accelerations $F_i$, dilatations $D$ and expansions $K$
do not preserve our EOM (\ref{fournine}).

The generators of the symmetries that remain  can be easily constructed.
The Hamiltonian $H$ can be read off from (\ref{fourtwo})
\begin{equation}\label{fourten}
H\,=\,(P_i\,+\,\frac{B}{2}\,\epsilon_{ij}\,x_j)x_i\,-\,E_ix_i.
\end{equation}
The space translation generators ${\cal P}_i$ are obtained by integrating
(\ref{fourfive})
\begin{equation}\label{foureleven}
{\cal P}_i\,=\,P_i\,-\,\frac{B}{2}\,\epsilon_{ij}\,x_j\,-\,E_it.
\end{equation}

Angular momentum generator $J$ is obtained from (\ref{threeseven})
by adding the term generating rotations of the $E_i$ fields
\begin{equation}\label{fourtwelve}
J\,=\,\epsilon_{ij}\, x_i\,P_j\,-\,\frac{\theta}{2}\,y_i^2\,+\,\epsilon_{ij}E_i\pi_j.
\end{equation}
The boost generators $K_i$ are constructed in close analogy with the corresponding expression in \cite{ho}
\begin{equation}\label{fourthirteen}
K_i\,=\,({\cal P}_i+\frac{E_i}{2}t)t\,+\,B\epsilon_{ij}\pi_j\,+\,\theta \epsilon_{ij}y_j.
\end{equation}
The operators $J_{\pm}$ are obtained from (\ref{threeten}) by adding
to it the term generating the required transformations  (47-48) of the electric fields
\begin{equation}\label{fourfourteen}
J_{\pm}\,=\,-\frac{\theta}{4}y_{\pm}^2\,\mp\,\frac{i}{2}(x_{\pm}P_{\pm}-E_{\pm}\pi_{\pm}).
\end{equation}
Using the EOM (40-43) one sees immediately that all the generators
(49-53) are conserved. Furthermore, using the PBs (34)
and (\ref{fourthree}) it can be shown that they generate the desired
transformations.

The enlarged Galilean symmetry algebra follows by the use of the fundamental PBs.
We find the following changes of the (2+1) dimensional exotic Galilean algebra:
\begin{equation}\label{fourfifteen}
\{P_i,\,H\}\,=\,0\quad \rightarrow\qquad \{{\cal P}_i,\,H\}\,=\,E_i
\end{equation}
\begin{equation}\label{foursixteen}
\{P_i,\,P_j\}\,=\,0\quad \rightarrow \qquad \{{\cal P}_i,\,{\cal P}_j\}\,=\,-B\epsilon_{ij}.
\end{equation}
The PBs (54) imply additional relations closing the algebra
with respect to $E_i$
\begin{equation}\label{fourseventeen}
 \{J,\,E_i\}\,=\,\epsilon_{ik}E_k,\qquad
 \{E_i,\,K_j\}\,=\,-B\epsilon_{ij},
\end{equation}
\begin{equation}\label{fournineteena}
 \{J_+,\,E_-\}\,=\,-iE_+,
\qquad
 \{J_-,\,E_+\}\,=\,iE_-.
\end{equation}
We note that $B$ plays the role of a new central element in the exotic
Galilean symmetry algebra, enlarged by the generators $J_{\pm}$ and $E_i$.
This 10-dimensional extended (2+1) dimensional Galilean algebra with two
central charges ($\theta,B$) describes the symmetries of the model (38).

 \section{Conclusions}
Scale and conformal transformations in nonrelativistic models were
first introduced in a way which  preserves the Schr\"odinger equation [3-5].
Such a procedure permits to add, in any dimensions, only two generators
$D$ and $K$ forming together with the Hamiltonian $H$ an $O(2,1)$ algebra of one dimensional
conformal transformations of the time variable (see also [9]).

 In this paper we have considered other
extensions of the Galilean nonrelativistic transformations by adding to them
conformal translations in the target space $x_i$ which describe the
constant accelerations. This has given us a nonrelativistic
conformal symmetry which can be shown to be derivable as the
nonrelativistic, {\it i.e.} $c\rightarrow \infty$ limit of the relativistic
conformal symmetry. We obtain in the relativistic and nonrelativistic cases
the same number of generators
and the requirement that the mass parameter vanishes.

Following our earlier considerations [1,2] we have considered in this
context  the nonrelativistic
dynamics in (2+1) dimensions, with the Galilean algebra endowed with two
central charges $m$ and $\theta$. We have found that our `genuine'
conformal extension of the Galilean symmetry requires $m=0$ and exhibits
the symmetries of the model recently considered in \cite{peterwojtek}.
It seems that the second central extension parameter $\theta\ne0$ is necessarily required if we wish to
construct a nonrelativistic conformally invariant free model.
We see that for the Galilean conformal
symmetry the parameter $\theta$ plays the analogous role  to the mass $m$ of standard Galilean
symmetries: it permits the explicit dynamical realisation of the symmetry
algebra. As the exotic central extension is  possible only in $D=(2+1)$ it is interesting to ask whether
 only in this dimension the Galilean conformal transformations
may represent an invariance group of a dynamical model.

The Galilean conformal symmetries can be also supersymmetrised by a suitable
contraction of the relativistic  superconformal algebras. In particular, in $D=(2+1)$,
such a structure is obtained by the contraction of the well known $OSp(N;4)$
superalgebra. The supersymmetric extension of the Galilean conformal
algebra is different from the `so-called' Schr\"odinger superalgebra
(see {\it e.g.} [11]) and is currently
under consideration.

\subsection*{Acknowledgments}
Two of the authors (JL) and (PCS) would like to thank the
University of Durham for hospitality and the EPSRC for financial support.


\begin{thebibliography}{99}

\bibitem{peterwojtek} P.C. Stichel and W.J. Zakrzewski, {\it Ann. Phys.}
{\bf 310}, 158 (2004).

\bibitem{lsz} J. Lukierski, P.C. Stichel and W.J. Zakrzewski,
 {\it Ann. Phys.} {\bf 260}, 224 (1997).


\bibitem{Hagen} C.R. Hagen, {\it Phys. Rev.} {\bf D5}, 377 (1972)

\bibitem{Niederev} U. Niederev, {\it Helv. Phys. Acta} {\bf 45}, 802 (1972)

\bibitem{Burdet} G. Burdet and M. Perrin, {\it Lett. Nuovo Cimento} {\bf 4}, 651 (1972)

\bibitem{Negro} J. Negro, M.A. del Olmo and A. Rodriguez-Mareo,
{\it J. Math. Phys.} {\bf 38}, 3786 (1997):
ibid. 3810 (1997).

\bibitem{ho} P.A. Horvathy, L. Martina and P.C. Stichel, {\it Phys. Lett.}
{\bf B615}, 87 (2005)

\bibitem{mack} G. Mack and A. Salam, {\it Ann. Phys.} {\bf 53}, 174 (1969)

\bibitem{fub} V. de Alfaro, S. Fubini and G. Furlan,
{\it Nuovo Cim.} {\bf 34A}, 569 (1978).

\bibitem{jackiw} L. Faddeev and R. Jackiw, {\it Phys. Rev. Lett.}
{\bf 60}, 1692 (1988).

\bibitem{du} C. Duval and P.A. Horvathy, {\it J. Maths. Phys} {\bf 35},
2516 (1994).


\end{thebibliography}
\end{document}